\documentclass[twocolumn,prl]{revtex4}
\usepackage{graphicx}
\usepackage{amsmath}

\begin{document}
\title{Is the Adiabatic Approximation Inconsistent?}
\author{Solomon Duki$^1$}
\author{H. Mathur$^1$}
\author{Onuttom Narayan$^2$}
\affiliation{$^1$Department of Physics, Case Western Reserve University, 
10900 Euclid Avenue, Cleveland OH 44106-7079 \\
$^2$Department of Physics, University of California, 
Santa Cruz, CA 95064}

\begin{abstract}

In recent Letters Marzlin and Sanders \cite{marzlin} and Tong 
{\em et al.} \cite{tong} study an adiabatically varying Hamiltonian
$h(t)$ that generates the time evolution $U(t)$ and its dual
$H(t)$ that generates the evolution $U^{\dagger}(t)$. Marzlin 
and Sanders show that inconsistent results are obtained if an 
adiabatic approximation is used to calculate $H(t)$; Tong 
{\em et al.} show that the adiabatic approximation can be very
inaccurate when applied to the exact dual Hamiltonian $H(t)$
even if it is an excellent approximation for $h(t)$. We show that
these two observations are equivalent and are not inconsistent with
the adiabatic theorem because in general, even if $h(t)$ satisfies 
the conditions of the adiabatic theorem, $H(t)$ will likely violate
those conditions.

\end{abstract}

\maketitle

The adiabatic theorem is the basis of an approximation
scheme that was discovered at the dawn of quantum mechanics \cite{born}
and that has been in widespread and continuous use ever since. 
Applications range from two-level systems (such as nuclei 
undergoing magnetic resonance or atoms interacting resonantly with
a laser field) to quantum field theory (where a low-energy effective
theory is derived by integrating out fast, high-energy degrees of
freedom). Two decades ago, Berry uncovered the beautiful geometric
structure underlying the adiabatic approximation \cite{berry},
leading to a resurgence of interest in the subject  and to new
applications \cite{shapere, anandan}.
More recently, it has been proposed that Berry phase effects
lead to quantum phase transitions that lie outside the usual
Landau-Ginzburg-Wilson paradigm \cite{senthil}.
The adiabatic theorem is also the basis of a newly proposed
quantum computing scheme \cite{farhi}.
Considering the significance of the adiabatic approximation to
quantum physics, the discovery of an inconsistency 
would be most disturbing. In a recent Letter Marzlin and Sanders 
\cite{marzlin} ask whether such an inconsistency might
exist, at least for a class of Hamiltonians. 
That question has been further studied by Tong {\em et al.} \cite{tong}
and subsequently commented on in Refs \cite{lidar,pati}. 
The purpose of this Letter is to show that there is no inconsistency.

Refs \cite{marzlin} and \cite{tong} start with a time-dependent
Hamiltonian $h(t)$ for which it is presumed that the adiabatic
approximation is accurate. The evolution operator for $h(t)$ is
denoted by $U(t),$ the solution to 
\begin{equation}
i \frac{\partial}{\partial t} U(t) = h(t) U(t),
\label{eq:uschrodinger}
\end{equation}
with $U(0) = {\cal I},$ where ${\cal I}$ is the
identity operator. Next, they
consider the dual Hamiltonian $H(t)$ for which the evolution
operator is $U^{\dagger}(t)$. Evidently
\begin{equation}
H(t) = - U^{\dagger}(t) i \frac{\partial}{\partial t} U(t) =
- U^{\dagger}(t) h(t) U(t).
\label{eq:dual}
\end{equation}
Tong {\em et al.} now formulate the putative inconsistency as follows:
First they argue that $H(t)$ satisfies the conditions for the adiabatic
theorem as well as does $h(t)$. Then they demonstrate that the
adiabatic approximation can be very inaccurate for the dual Hamiltonian.
These conflicting observations constitute the ``inconsistency''. 
Marzlin and Sanders \cite{marzlin} originally formulated the 
inconsistency in a different form; below we will show the equivalence of the
two formulations.

In this Letter we examine the dual Hamiltonian
$H(t)$ and find it violates well known adiabaticity conditions;
hence there is no inconsistency. The chief difficulty in determining
whether $H(t)$ satisfies adiabatic conditions is that we do not
have an explicit expression for $H(t)$ except in the special
cases where the dynamics of $h(t)$ are simple enough to allow
evaluation of $U(t)$. Nonetheless, we are able to give a general
argument that $H(t)$ violates the conditions of the adiabatic
theorem. As an illustration, we apply our general arguments
to a solvable two-level model also studied by ref \cite{tong}.
In this case it is possible to obtain the explicit form of
$H(t)$; inspection of this form is sufficient to show 
immediately that $H(t)$ violates the conditions of the adiabatic
theorem, consistent with our result.


It is helpful to first review the conditions under which
the adiabatic approximation is accurate. Consider
a two-level system with the Schr\"{o}dinger equation
\begin{equation}
i \frac{\partial}{\partial t} 
\left( 
\begin{array}{c}
c_1 \\
c_2 \\
\end{array}
\right) 
= \frac{1}{2}
\left( 
\begin{array}{cc}
\omega_0 & \Omega e^{-i \omega t} \\
\Omega e^{i \omega t} & - \omega_0 \\
\end{array}
\right)
\left(
\begin{array}{c}
c_1 \\
c_2 \\
\end{array}
\right).
\label{eq:twolevel}
\end{equation}
We regard the off-diagonal terms as a perturbation and
ask when the perturbation may be neglected. Evidently,
the perturbation must be small in magnitude, 
but even if 
it is,
it can have a big
effect on resonance when $\omega \approx \omega_0$. This is more
transparent if we go over to the interaction picture by 
writing $c_1 = a_1 \exp( - i \omega_0 t/2 )$, $c_2 = a_2
\exp( i \omega_0 t/2 )$. In the interaction picture
\begin{equation}
i \frac{\partial}{\partial t}
\left(
\begin{array}{c}
a_1 \\
a_2 \\
\end{array}
\right)
= 
\frac{1}{2} 
\left(
\begin{array}{cc}
0 & \Omega e^{-i(\omega - \omega_0)t} \\
\Omega e^{i (\omega - \omega_0) t} & 0 \\
\end{array}
\right)
\left( 
\begin{array}{c}
a_1 \\
a_2 \\
\end{array}
\right).
\label{eq:2levelinteraction}
\end{equation}
Away from resonance the off-diagonal terms oscillate rapidly
and average to zero. Near resonance, the perturbation varies slowly
in the interaction picture and can have a big effect,
producing Rabi oscillations. The precise condition to be 
off-resonance is $|\omega - \omega_0| \gg \Omega$.
For this problem, the condition can be derived
by transforming to another rotating frame via
$a_1 = b_1 \exp[ - i (\omega - \omega_0)t/2 ]$,
$a_2 = b_2 \exp[i (\omega - \omega_0)t/2 ]$. In this 
frame, the Hamiltonian is time independent and equal
to $(1/2)[ (\omega - \omega_0) \sigma_z + \Omega \sigma_x ]$.
In summary, we need the perturbation to be 
off-resonance ($\Omega \ll | \omega -
\omega_0 |$) for it to be truly negligible.

Now let us consider the problem of a general time dependent hamiltonian
$h(t).$ It is helpful to use a slightly different approach from the 
one above, which is useful for adiabatic perturbations. To this end, we
introduce the instantaneous
eigenstates $|n(t) \rangle$ that satisfy
\begin{equation}
h(t) |n(t) \rangle = \epsilon_n(t) |n(t) \rangle.
\label{eq:instestates}
\end{equation}
We choose the phases of $|n(t) \rangle$ to satisfy 
$ \langle n(t) | \partial/\partial t | n(t) \rangle = 0 $,
a convention called the parallel transport gauge \cite{anholonomy}. 
We expand the state of the system
$ | \psi \rangle $ in this time-dependent basis. Thus
\begin{equation}
| \psi (t) \rangle = 
\sum_{n} \phi_n (t) \exp[ - i \int_0^t d t' \varepsilon_n(t') ] 
| n(t) \rangle.
\label{eq:psin}
\end{equation}
In this moving frame the time-dependent Schr\"{o}dinger
equation has the form
\begin{equation}
i \frac{ \partial }{ \partial t } \phi_n (t) = 
\sum_{n \neq m} A_{nm} (t) 
\exp( i \int_{0}^{t} d t' [ \varepsilon_n(t') -
\varepsilon_m(t') ] ) \phi_n (t).
\label{eq:adiabaticinteraction}
\end{equation}
Here
\begin{equation}
A_{nm} (t) = -i \langle n | \frac{ \partial}{ \partial t} | m \rangle.
\label{eq:anm}
\end{equation}
The adiabatic approximation amounts to neglect of the terms 
on the right hand side of eq (\ref{eq:adiabaticinteraction}).
For this to be justified, by analogy to eq (\ref{eq:2levelinteraction}),
we see that the neglected terms must be off-resonance. Roughly
the condition to be off-resonance is that the neglected terms should
{\em not} vary slowly; i.e. the terms should be small in
magnitude compared to their predominant frequency. Note that this is
somewhat different from the adiabatic condition $ |A_{nm}| \ll 
| \varepsilon_n - \varepsilon_m | $ used by refs \cite{marzlin,tong}.
More precisely, if the Hamiltonian varies predominantly at a 
frequency $\omega,$ and the typical spacing between instantaneous 
eigenvalues is $\Delta,$ our adiabatic condition states 
$| A_{nm} | \ll (\Delta - \omega)$. Alternatively, if the Hamiltonian
varies on a time scale $T$ for which $1/\Delta T\ll 1,$ and 
$A_{nm}\sim 1/T,$ the correction to the adiabatic approximation is 
$\sim 1/\Delta T$ which is very small. This condition is more restrictive than
neccessary, but it agrees with the intuitive expectation 
that for the adiabatic approximation to apply, the Hamiltonian
must `` vary slowly''.



Sampling the literature we find that the graduate text by 
Schiff \cite{schiff} gives a rather complete discussion
of the adiabatic approximation, emphasizing that the neglected
terms must be non-resonant. On the other
hand, Landau and Lifshitz \cite{landau} give the more
restrictive adiabatic condition $T \rightarrow \infty$.
Moody, Shapere and
Wilczek \cite{moody} compute non-perturbative corrections
to adiabatic evolution using $1/\Delta T$ as the small
parameter. Berry \cite{iteration} also regards $1/\Delta T$
as the adiabatic parameter and suggests that the corrections
vanish as $\exp( - \Delta T )$, also proposed by
Hwang and Pechukas \cite{hwang}. Thus it appears to be generally
accepted that it is sufficient for the Hamiltonian to be slowly
varying, but less restrictive conditions are also discussed. 

For the dual Hamiltonian 
$H(t),$ denote the instantaneous eigenstates $ |n(t); H \rangle$
and the eigenvalues $\varepsilon_n^H (t)$. With the  parallel
transport gauge, in the adiabatic frame 
\begin{equation}
| \psi (t) \rangle = \sum_{n} \phi_n^H (t)
\exp[ - i \int_0^t d t' \varepsilon_n^H (t') ] | n(t); H \rangle.
\label{eq:Hansatz}
\end{equation}
By analogy to eq (\ref{eq:adiabaticinteraction}) the 
Schr\"{o}dinger equation obeyed by the amplitudes 
$\phi_n^H(t)$ is
\begin{equation}
i \frac{ \partial }{\partial t} \phi_n^H (t) = 
\sum_{m \neq n} A_{nm}^H (t) 
\exp( i \int_0^t d t' [\varepsilon_n^H (t') - \varepsilon_m^H (t')] )
\phi_m^H(t).
\label{eq:Hinteraction}
\end{equation}

Although, as noted above, we do not have an explicit expression for
$H(t)$, it is easy to relate the eigenstates of $H$ to those of $h$.
Evidently $U^{\dagger}(t) | n (t) \rangle$ is an eigenstate of $H(t)$
with eigenvalue $ - \varepsilon_n (t)$. Hence we write
\begin{eqnarray}
| n(t); H \rangle & = & U^{\dagger} (t) | n(t) \rangle
\exp[ - i \int_0^t d t' \varepsilon_n(t')] ;
\nonumber \\
\varepsilon_n^H (t) & = & - \varepsilon_n (t).
\label{eq:relate}
\end{eqnarray}
The phase of $|n(t); H \rangle$ has been chosen to
ensure parallel transport, $ \langle n; H | \partial_t | n; H \rangle 
= 0$. To verify this, it is helpful to recall that 
$i U \partial_t U^{\dagger} = - h,$
which follows from the Schr\"{o}dinger equation (\ref{eq:uschrodinger}).
Using eq (\ref{eq:relate}) and this result, one can show that 
\begin{equation}
A_{nm}^H (t) = A_{nm}(t) \exp( i \int_0^t d t'
[ \varepsilon_m (t') - \varepsilon_n (t') ]).
\label{eq:arelation}
\end{equation}
Substituting eq (\ref{eq:arelation}) into eq (\ref{eq:Hinteraction}) we obtain
the final form of the Schr\"{o}dinger equation in the adiabatic frame for the
dual Hamiltonian $H(t)$,
\begin{equation}
 i \frac{\partial}{\partial t} \phi_n^H (t) =
\sum_{m \neq n} A_{nm} (t) \phi_m^H (t).
\label{eq:clearview}
\end{equation}
Eq (\ref{eq:clearview}) is 
the main result of our general analysis.

Eq (\ref{eq:clearview}) shows that the terms that would be neglected in the
adiabatic approximation $A_{nm}(t)$ vary slowly. The typical
frequency of these terms is $1/T$, the same as their magnitude.
Thus $H(t)$ does not fulfil the adiabatic condition, 
and using the adiabatic approximation for $H(t)$ leads to inaccurate results. 

This concludes our general analysis of $H(t)$. We turn to a 
solvable example. Take the two-level Hamiltonian
\begin{equation}
h(t) = - \frac{1}{2} \omega_0
\left( 
\begin{array}{cc}
\cos \theta & \sin \theta e^{-i \omega t} \\
\sin \theta e^{i \omega t} & - \cos \theta \\
\end{array}
\right)
\label{eq:solvableh}
\end{equation}
Physically we can picture this as a spin $\frac{1}{2}$ particle in a magnetic
field tilted at an angle $\theta$ to the $z$-axis and rotating at a frequency
$\omega$. Essentially this model was also studied in ref \cite{tong}. The
instantaneous eigenvalues of this Hamiltonian are $ \varepsilon_{\pm} (t) =
\pm \omega_0/2$; the corresponding eigenspinors in the parallel transport
gauge are \cite{notetwo}
\begin{eqnarray}
|+(t) \rangle & = & 
\left(
\begin{array}{c}
\cos \frac{ \theta }{2} \\
\sin \frac{\theta}{2} e^{i \omega t} \\
\end{array}
\right) 
\exp[ - i \frac{ \omega t}{2} (1 - \cos \theta)],
\nonumber \\
|-(t) \rangle & = &
\left(
\begin{array}{c} 
- \sin \frac{\theta}{2} e^{- i \omega t} \\
\cos \frac{\theta}{2} \\
\end{array}
\right)
\exp[ i \frac{\omega t}{2} (1 - \cos \theta)].
\label{eq:hspinors}
\end{eqnarray}
A straightforward computation reveals that
\begin{equation}
A_{+ -} (t) = 
\langle + | \partial_t | - \rangle =
\frac{ \omega }{2} \sin \theta 
e^{- i \omega t \cos \theta }.
\label{eq:aplusminus}
\end{equation}

It follows from eq (\ref{eq:adiabaticinteraction}) that the 
adiabatic frame 
Schr\"{o}dinger equation for $h(t)$ is 
\begin{equation}
i \frac{ \partial }{ \partial t } 
\left(
\begin{array}{c} 
\phi_+ \\
\phi_- \\
\end{array}
\right) = \left(
\begin{array}{cc}
0 & A_{+-} e^{ i \omega_0 t }\\
A_{+-}^* e^{- i \omega_0 t} & 0 \\
\end{array}
\right)
\left(
\begin{array}{c}
\phi_+ \\
\phi_- \\
\end{array}
\right).
\label{eq:hinteraction}
\end{equation}
Clearly for the off-diagonal term to be off-resonance,
we need $| \omega \cos \theta - \omega_0 | \gg \omega \sin \theta$. 
Evidently, this condition is satisfied if $ \omega \ll \omega_0 $, {\em 
i.e.}, $h(t)$ varies slowly.

It follows from eq (\ref{eq:clearview}) that the adiabatic frame
Schr\"{o}dinger equation for $H(t)$ is
\begin{equation}
i \frac{\partial}{\partial t} 
\left(
\begin{array}{c}
\phi_+ \\
\phi_- \\
\end{array}
\right) =
\left( 
\begin{array}{cc}
0 & A_{+-}(t) \\
A_{+-}^*(t) & 0 \\
\end{array}
\right) 
\left( 
\begin{array}{c}
\phi_+ \\
\phi_- \\
\end{array}
\right).
\label{eq:Hinteractiontwo}
\end{equation}
Clearly, for the off-diagonal term to be off-resonance
we need $ | \omega \sin \theta | \ll | \omega \cos \theta|$, a 
condition met only when $\theta \ll 1$ (or $ \pi - \theta \ll 1$). 

In summary we find that for $h(t)$ to be adiabatic it is sufficient
that $\omega \ll \omega_0$; but for $H(t)$ to be adiabatic we also need
$ \theta \ll 1 $. In this limit, $\theta \rightarrow 0$, 
the fidelity of the adiabatic solution to $H(t)$ (the overlap of the 
adiabatic and exact solutions) computed by 
\cite{marzlin,tong} approaches unity, consistent with our finding.

We note that in this solvable problem it is possible to 
compute $U$ and explicitly obtain 
$H(t)$. This column is far too narrow to write the 
entire expression, but it includes
terms that oscillate at a frequency 
\begin{equation}
\nu = \sqrt{\omega_0^2 + \omega^2 + 2 \omega_0 \omega \cos \theta}.
\label{eq:nu}
\end{equation}
In the limit $\omega \ll \omega_0$, needed for the Hamiltonian $h(t)$
to be adiabatically varying, $ \nu \rightarrow \omega_0$. Thus, even 
without going to to the adiabatic frame, a cursory
inspection of $H(t)$ is sufficient to show it is not slowly varying and
is unlikely to satisfy the adiabatic condition.  

We briefly comment on cases where the adiabatic approximation 
applies to $h(t)$ and $H(t).$ This happens when there is a parameter other than
$T$ which can be tuned to make the magnitude of the 
off-diagonal term in eq (\ref{eq:clearview}) small 
compared to the predominant frequency $1/T$. 
However, for $h(t)$ the approximation 
becomes more accurate as $T \rightarrow 
\infty$ (with exponentially small corrections according to
refs \cite{moody,iteration,hwang}), whereas for for $H(t)$ the
distance from resonance is essentially independent of $T$ 
and is controlled by the additional parameter.
This can be seen in the example above, 
where $T \rightarrow 2 \pi/\omega$
and $\theta$ is the additional parameter. 

We make a few observations in passing here about the
discussion of the adiabatic conditions after Eq.(\ref{eq:anm}). 
First, the effects
of $A_{nm}$ are implicitly integrated over a finite
time window. This is appropriate, since in experiments
$h(t)$ is typically varied only within a finite time
window. However, it is the discontinuity in $dh/dt$ that
causes the $\sim 1/(\Delta T)$ correction to the adiabatic
approximation for large $T;$
if all derivatives of $h(t)$ are continuous for all $t,$
it is known \cite{moody,iteration,hwang} that the correction will
be exponentially small in $T\Delta.$ Second, $\Delta$
was effectively taken to be time independent. When both 
$A_{nm}$ and $\Delta$ vary sinusoidally at frequency $\omega,$
higher order resonances result when $\omega = \Delta_0/k$
for integer $k.$ This is essentially the same mechanism
that causes resonances going beyond first order
perturbation theory. 

Finally we discuss the equivalence between the formulations
of refs \cite{marzlin} and \cite{tong}.  
They assume that the adiabatic approximation is accurate for 
$h(t):$ $U(t) \approx U_{{\rm adia}}(t)$
where
\begin{equation}
U_{{\rm adia}} (t) = \sum_{n} |n(t) \rangle \langle n(0)| 
\exp \left[ - i \int_{0}^{t} d t' \varepsilon_n(t') \right]
\label{eq:uadia}
\end{equation}is the adiabatic approximation to the exact evolution
$U(t)$. Marzlin {\em et al.} \cite{marzlin} then develop an 
approximation to $U^{\dagger}(t)$ that we denote $V^{\dagger}(t)$.
They approximate the dual Hamiltonian [defined by eq (\ref{eq:dual})]
as $H^{(1)}_{{\rm adia}} (t)  =  - U^{\dagger}_{{\rm adia}} (t) h(t) 
U_{{\rm adia}} (t).$
They compute the evolution $V^{\dagger}(t)$ corresponding to
$H^{(1)}_{{\rm adia}}(t),$ obtaining 
\begin{equation}
V^{\dagger}(t) = \sum_{n} | n(0) \rangle \langle n(0) | 
\exp \left[ i \int_{0}^{t} d t' \varepsilon_n(t') \right].
\label{eq:vdagger}
\end{equation}
In contrast, Tong {\em et al.} compute a different approximation to
$U^{\dagger} (t)$ that we denote $W^{\dagger}(t)$. They work with the
exact dual Hamiltonian $H(t)$, but work out its evolution using the
adiabatic approximation. Analogy to eq (\ref{eq:uadia}) and use of 
eq (\ref{eq:relate}) leads to the result
\begin{equation}
W^{\dagger}(t) = \sum_{n} U^{\dagger} (t) | n(t) \rangle \langle n(0) |.
\label{eq:wdagger}
\end{equation}
The operators $V^{\dagger}$ and $W^{\dagger}$ are different
in appearance and in the approximations that lead to them, but we will
show they are equivalent to the extent that the adiabatic 
approximation $U \approx U_{{\rm adia}}$ is valid. 

With this notation established we now turn to the inconsistencies.
Marzlin and Sanders \cite{marzlin} consider the identity $U U^{\dagger}
= {\cal I}$ and replace $U \rightarrow U_{{\rm adia}}$ and 
$U^{\dagger} \rightarrow V^{\dagger}$ with the disastrous result that
$U U^{\dagger} \rightarrow \sum_{n} | n (t) \rangle \langle n(0) | 
\neq {\cal I}.$
Tong {\em et al.} derive the same inconsistency by replacing $U^{\dagger}$
with $W^{\dagger}:$   
$U U^{\dagger} \rightarrow U W^\dagger$ and using Eq.(\ref{eq:wdagger}).
The equivalence of the two approaches can be seen by starting with the
trivial identity $U^{\dagger} = U^{\dagger} U U^{\dagger}$. If on the 
right hand side we replace $U^{\dagger} U U^{\dagger} \rightarrow 
U^{\dagger} U_{{\rm adia}} V^{\dagger}$ we obtain $W^{\dagger}$ by
use of eqs (\ref{eq:uadia}), (\ref{eq:vdagger}) and (\ref{eq:wdagger}).
In other words, the approximation of Marzlin and Sanders 
with the adiabatic approximation is equivalent to the approximation
of Tong {\em et al.}

The resolution of the inconsistencies of Refs.\cite{marzlin,tong}
is that $V^{\dagger} \approx U^{\dagger}$ and
$W^{\dagger} \approx U^{\dagger}$ are not good approximations; the
adiabatic approximation $U_{{\rm adia}} \approx U$ is not at fault.
In this Letter we have explained the failure of the approximation
$W^{\dagger} \approx U^{\dagger}$. This may also be considered a
resolution of the Marzlin and Sanders form of the inconsistency,
due to the equivalence shown above. An 
alternative resolution was provided in ref \cite{marzlin} who used
the adiabatic approximation to calculate a second approximation to
the dual Hamiltonian 
$H^{(2)}_{{\rm adia}} (t)  = - i U^{\dagger}_{{\rm adia}}
\partial_t U_{{\rm adia}}.$
By construction, the evolution operator corresponding to 
$H^{(2)}_{{\rm adia}}(t)$ is $U^{\dagger}_{{\rm adia}}(t)$ 
which is indeed a good approximation to $U^{\dagger}$. Since the
trustworthy
adiabatic approximation has been used to compute both
$H^{(1)}_{{\rm adia}}(t)$ and $H^{(2)}_{{\rm adia}}(t)$ 
we may presume that $H^{(1)}_{{\rm adia}}(t) \approx 
H^{(2)}_{{\rm adia}} (t) \approx H(t)$. This does not mean that
the three Hamiltonians will generate essentially the same evolution
since small errors can grow upon time exponentiation. Indeed 
ref~\cite{marzlin} finds that $H^{(1)}_{{\rm adia}}$ and
$H^{(2)}_{{\rm adia}}$ differ by a small resonant perturbation
leading to the inference that $V^{\dagger}$ may be very different
from $U^{\dagger}_{{\rm adia}}$. Since $U^{\dagger}_{{\rm adia}}
\approx U^{\dagger}$ this may be considered to explain the
failure of the approximation $V^{\dagger} \approx U^{\dagger}.$


To summarise, we have studied the dual pair of 
Hamiltonians $h(t)$ and $H(t)$ that generate time evolution
$U(t)$ and $U^{\dagger} (t)$ respectively. 
Marzlin and Sanders \cite{marzlin} showed that if an adiabatic
approximation is used to compute $H(t)$ an inconsistency 
results. Tong {\em et al.} \cite{tong}
showed that the same inconsistency results if the adiabatic
approximation is applied to the exact dual Hamiltonian $H(t)$.
We show that these observations are essentially equivalent.
Our main finding
is that even if $h(t)$ satisfies the conditions of the
adiabatic theorem, $H(t)$ will not (except in the trivial case when 
the total change in $h(t)$ is small) because the terms 
neglected in the adiabatic approximation are resonant for $H(t).$
Thus 
the breakdown of 
the adiabatic approximation for $H$ is not inconsistent
with the adiabatic theorem.

It is a pleasure to acknowledge helpful discussions with
Francesc Ferrer.


\begin{thebibliography}{99}

\bibitem{born} M. Born and V. Fock, Z Physik {\bf 51}, 165 (1928).

\bibitem{berry} M.V. Berry, Proc. Roy. Soc. London {\bf A392}, 45
(1984).

\bibitem{shapere} For a collection of early reprints with valuable 
commentary, see A. Shapere and F. Wilczek, {\em Geometric Phases 
in Physics} (World Scientific, Singapore, 1989).

\bibitem{anandan} For a bibliography see {\em e.g.}  J. Anandan,
J. Christian and K. Wanelik, {\em Resource Letter GPP-1:
Geometric Phases in Physics}, Am J Phys {\bf 65}, 180 (1997).

\bibitem{senthil} For an introduction, see S. Sachdev, 
{\em Quantum Phase Transitions} (Cambridge University Press, 1999). 
For a recent review, see T. Senthil, cond-mat/0411275.

\bibitem{farhi} E. Farhi {\em et al.}, Science {\bf 292}, 472 (2001).

\bibitem{marzlin} K.-P. Marzlin and B.C. Sanders, Phys. Rev. Lett.
{\bf 93}, 160408 (2004).

\bibitem{tong} D.M. Tong, K. Singh, L.C. Kweh and C. H. Oh, 
Phys. Rev. Lett. {\bf 95}, 110407 (2005).

\bibitem{lidar} M.S. Sarandy, L.-A. Wu and D.A. Lidar,
Quant. Info. Proc. {\bf 3}, 331 (2004); quant-ph/040509v3.

\bibitem{pati} A.K. Pati and A.K. Rajagopal, quant-ph/0405129v1.

\bibitem{anholonomy} Suppose $h(t_2) = h(t_1)$. With our phase 
convention $ |n(t_2) \rangle = |n(t_1) \rangle \exp( - i \gamma )$
where $\gamma $ is Berry's phase for adiabatic evolution between 
$t_1$ and $t_2$. An explicit example will be discussed below.

\bibitem{schiff} L.I. Schiff, {\em Quantum Mechanics} (McGraw-Hill, New 
York, 1955).

\bibitem{landau} L.D. Landau and E.M. Lifshitz, {\em Course of Theoretical
Physics, vol 3, Non-Relativistic Quantum Mechanics} (Pergamon Press,
Oxford, 1977).

\bibitem{moody} See the article by 
J. Moody, A. Shapere and F. Wilczek in ref \cite{shapere},
p 160-183.

\bibitem{iteration} M.V. Berry, Proc. Roy. Soc. London, {\bf A414}, 31 (1987),
reprinted in \cite{shapere}.

\bibitem{hwang} J.-T. Hwang and P. Pechukas, J. Chem. Phys. {\bf 67}, 
4640 (1977).

\bibitem{notetwo} For the benefit of readers of endnote \cite{anholonomy}
we observe that the Hamiltonian in eq (\ref{eq:solvableh}) is 
periodic under $ t \rightarrow t + 2 \pi/\omega$. Under this time
translation, the instantaneous eigenstates, eq (\ref{eq:hspinors}),
change according to $| \pm \rangle \rightarrow
| \pm \rangle \exp( \mp i \gamma )$. Here
$\gamma = \pi (1 - \cos \theta)$ is one half the
solid angle traced by the magnetic field, the well-known
Berry phase for this problem \cite{berry}.

\end{thebibliography}
\end{document}